\documentclass[twocolumn,secnumarabic,amssymb, nobibnotes, aps, prd]{revtex4-1}

\newcommand{\bapp}{\mathrm{B}_\mathrm{app}}
\newcommand{\bexch}{\mathrm{B}_\mathrm{exch}}
\newcommand{\vhbn}{\mathrm{V_{hBN}}}
\newcommand{\pin}{\mathrm{P}_\mathrm{I}}
\newcommand{\pd}{\mathrm{P}_\mathrm{D}}

\newcommand{\rnl}{\mathrm{R}_\mathrm{NL}}
\newcommand{\rsq}{\mathrm{R}_\mathrm{sq}}

\newcommand{\gc}{^\circ\mathrm{C}}

\newcommand{\ds}{\mathrm{D}_{\mathrm{s}}}

\newcommand{\ts}{\tau_{\mathrm{s}}}

\newcommand{\ohm}{\Omega}

\usepackage{graphicx}
\usepackage{color}

\usepackage{amsmath}
\usepackage{dcolumn}
\usepackage{bm}
\usepackage{float}

\begin{document}

\title{Bias dependent spin injection into graphene on YIG through bilayer hBN tunnel barriers}%

\author{J.C. Leutenantsmeyer}
\email[]{E-Mail: j.c.leutenantsmeyer@rug.nl}
\affiliation{Physics of Nanodevices, Zernike Institute for Advanced Materials, University of Groningen, 9747 AG Groningen, The Netherlands}

\author{T. Liu}
\affiliation{Physics of Nanodevices, Zernike Institute for Advanced Materials, University of Groningen, 9747 AG Groningen, The Netherlands}

\author{M. Gurram}
\affiliation{Physics of Nanodevices, Zernike Institute for Advanced Materials, University of Groningen, 9747 AG Groningen, The Netherlands}

\author{A.A. Kaverzin}
\affiliation{Physics of Nanodevices, Zernike Institute for Advanced Materials, University of Groningen, 9747 AG Groningen, The Netherlands}

\author{B.J. van Wees}
\affiliation{Physics of Nanodevices, Zernike Institute for Advanced Materials, University of Groningen, 9747 AG Groningen, The Netherlands}

\begin{abstract}
We study the spin injection efficiency into single and bilayer graphene on the ferrimagnetic insulator Yttrium-Iron-Garnet (YIG) through an exfoliated tunnel barrier of bilayer hexagonal boron nitride (hBN). The contacts of two samples yield a resistance-area product between 5 and 30~k$\ohm\mu$m$^2$. Depending on an applied DC bias current, the magnitude of the non-local spin signal can be increased or suppressed below the noise level. The spin injection efficiency reaches values from -60\% to +25\%. The results are confirmed with both spin valve and spin precession measurements. The proximity induced exchange field is found in sample~A to be (85 $\pm$ 30)~mT and in sample~B close to the detection limit. 
Our results show that the exceptional spin injection properties of bilayer hBN tunnel barriers reported by Gurram et al. are not limited to fully encapsulated graphene systems but are also valid in graphene/YIG devices. This further emphasizes the versatility of bilayer hBN as an efficient and reliable tunnel barrier for graphene spintronics.
\end{abstract}

\date{\today}%
\maketitle

\date{\today}%
\maketitle

\section{Introduction}
The combination of graphene with other two dimensional layered materials is an elegant way to create atomically thin devices with adjustable properties \cite{Geim2013,Han2014,Roche2015}. The crystalline insulator hexagonal boron nitride is an appealing material for the field of graphene spintronics \cite{Gurram2018}. Its atomic flatness and sufficiently strong van der Waals interaction with graphene allows the fabrication of heterostructures of 2D materials with minimized contamination, implying good spin transport properties. A long spin diffusion length of 30 $\mu$m has been experimentally achieved in graphene where a bulk flake of hBN was used as protective layer to avoid contamination during the fabrication process \cite{Drogeler2016}. Therefore, the use of hBN as a pinhole free tunnel barrier is straightforward since these fully encapsulated graphene devices suggest minimized contamination and highly efficient spin transport. Several experimental studies have investigated the spin injection through tunnel barriers of exfoliated hBN \cite{Yamaguchi2016,Gurram2016} and large scale hBN grown via chemical vapor deposition \cite{Fu2014,Kamalakar2015,Kamalakar2016,Gurram2018a}. However, the experimentally demonstrated spin transport lengths are still far below the values suggested by the low intrinsic spin orbit coupling of graphene \cite{Huertas-Hernando2006}. 

Having graphene in proximity to magnetic materials is a novel approach to tune the intrinsic properties of graphene. Magnetic graphene is characterized by the induced exchange field \cite{Leutenantsmeyer2017,Singh2017,Wei2016,Wang2015,Asshoff2017}. First principle calculations of idealized systems predict an exchange splitting of the graphene spin states to exceed several tens of meV \cite{Yang2013,Hallal2017}. However, the experimentally demonstrated exchange fields are still several orders of magnitude below \cite{Leutenantsmeyer2017,Singh2017,Evelt2016}.

The realization of graphene devices with a large exchange field requires the tackling of several challenges. The cleanliness of the interface between graphene and YIG is crucial to obtain a strong exchange effect as indicated by the discrepancy between experimentally achieved values and theoretical predictions. Furthermore, the interface and tunnel barrier between the graphene flake and contacts are crucial for the injection of a large spin accumulation and the observation of large spin signals. In our previous works we employed tunnel barriers of oxidized titanium or aluminum to overcome the conductivity mismatch problem \cite{Schmidt2000,Rashba2000}. For these types of tunnel barrier the magnitude of the spin signal is limited by pinholes and resulted in a relatively small spin signal of mostly less than 1~$\ohm$, which often did not exceed the electrical noise of the measured signals in the sample. In addition, the contamination arising from the PMMA-based fabrication procedure affects the graphene cleanliness negatively. 
For this study we replace the AlO$_\mathrm{x}$ or TiO$_\mathrm{x}$ tunnel barrier with a bilayer-hBN (bl-hBN) flake, which significantly improves the sample quality and spin signal. Furthermore, we confirm the tunable spin injection reported by Gurram et al. \cite{Gurram2017} for the graphene/YIG system.

\section{Sample preparation and contact characterization}
Thin hBN flakes are exfoliated from hBN crystals (HQ Graphene) onto 90~nm SiO$_2$ wafers. The thickness of the flakes is estimated through their optical contrast, which is calibrated by atomic force microscopy. In our microscope (Zeiss Axio Imager.A2m with an EC Epiplan-Neofluar 100x/0.9 objective) bl-hBN corresponds to 2.5\% contrast in the green channel. Suitable bl-hBN flakes are picked up by using a dry polycarbonate based transfer method \cite{Zomer2014} and combined with single- (sample~A) or bilayer graphene (sample~B) exfoliated from HOPG crystals (ZYB grade, HQ Graphene). The stack is placed on a cleaned 12~$\mu$m YIG grown by liquid phase epitaxy (LPE) on a 600~$\mu$m gadolinium-gallium-garnet substrate (Matesy GmbH). Before the transfer, the YIG substrate for sample~A is treated with oxygen plasma to remove organic contaminants and annealed in a 500$\gc$ furnace in an oxygen atmosphere prior to the transfer of the graphene/bl-hBN stack. The substrate of sample~B underwent an additional argon plasma treatment before the annealing step. 

\begin{figure}[htb]
\centerline{\includegraphics[width=1\linewidth]{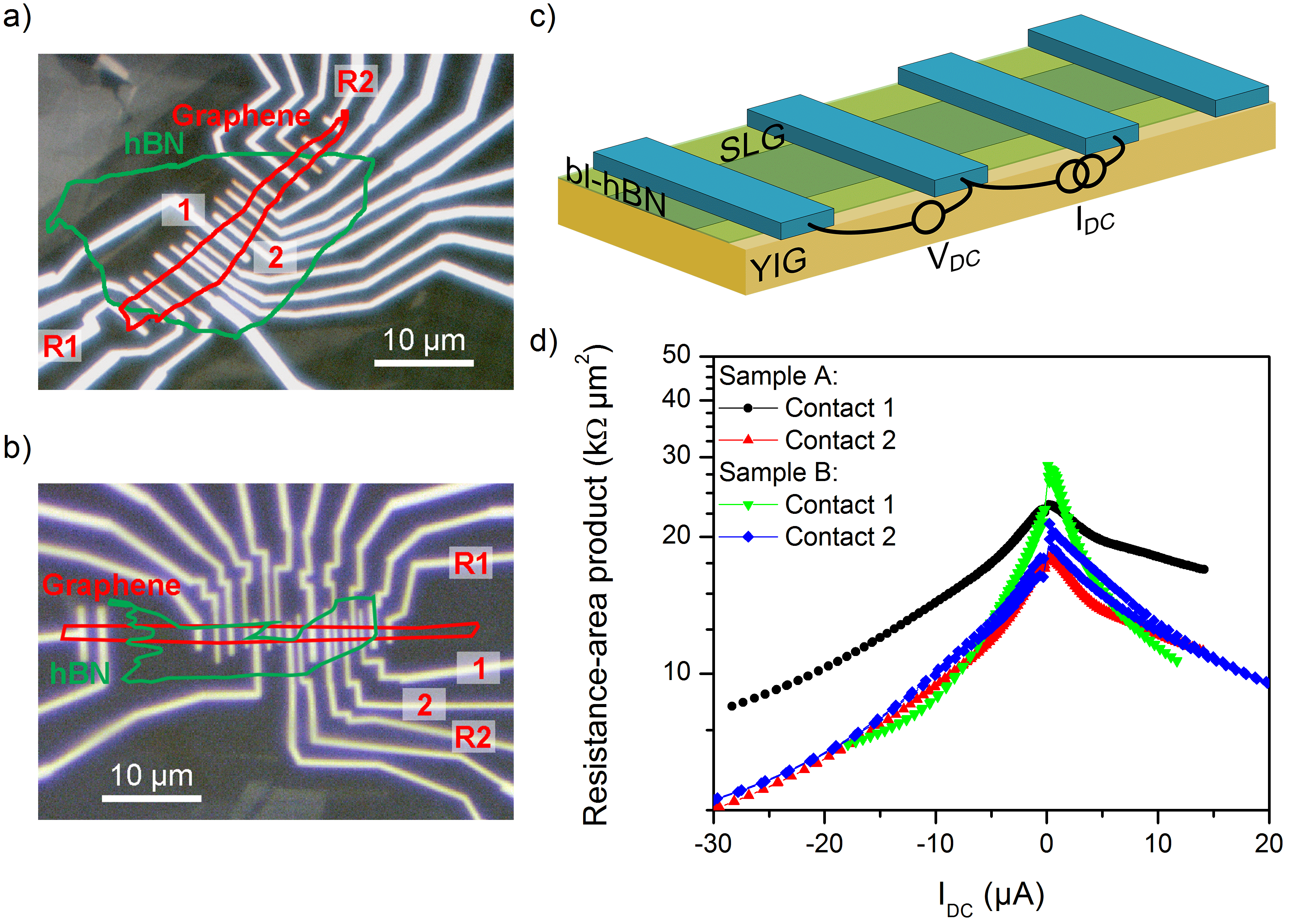}}
\caption{a) Optical micrograph of the sample~A. The outer electrodes (R) are not covered by bl-hBN and used as reference electrodes in both local and non-local measurements. b) Optical micrograph of sample~B. c) Schematic measurement of the three-terminal contact resistance.
d) All working contacts have a calculated resistance-area product between 5 and 30~k$\ohm\mu$m$^2$. The full set of IV characteristics is shown in the supplementary information.
\label{Fig1}}
\end{figure}

The polycarbonate is dissolved in chloroform and the bl-hBN/graphene/YIG stack is cleaned in acetone, isopropanol and sequent annealing for one hour at 350$\gc$ in an argon-hydrogen atmosphere. Contacts are defined using a standard PMMA-based electron beam lithography process. The electrodes are evaporated at pressures below 10$^{-7}$~mbar and consist of 45~nm cobalt and a 5~nm aluminum capping layer. After the liftoff in warm acetone, the sample (Figs.~\ref{Fig1}a and \ref{Fig1}b) is loaded into a cryostat and kept in vacuum during the characterization. All measurements are carried out at 75~K.

After loading into the cryostat of the measurement setup, the samples are cooled down to liquid nitrogen temperature and the contacts are characterized in a three-terminal geometry (Fig.~\ref{Fig1}c) using the outermost contacts as reference electrodes. The resistance-area product is calculated from the current-voltage characteristics and shown for sample~A in Fig.~\ref{Fig1}d. The contacts on sample~A and B which employ a bl-hBN tunnel barrier yield a typical resistance-area product between 5 and 30~k$\ohm\mu$m$^2$, a range comparable to the one reported in \cite{Gurram2017}. An hBN covered graphene Hall bar sample fabricated in parallel with sample~B for comparison yields a carrier density of n = $5\times 10^{12}$~cm$^{-2}$ and a mobility of $\mu = 5400\,\mathrm{cm}^2$/Vs. We found $\mu = 720\,\mathrm{cm}^2$/Vs (estimated via the Shubnikov-de Haas oscillations) in our previous work \cite{Leutenantsmeyer2017} and conclude that the protective hBN layer significantly improves the graphene charge transport properties on YIG. 

\section{Bias-dependent spin injection through bilayer hBN tunnel barriers into single and bilayer graphene on YIG}
\begin{figure}[htb]
\centerline{\includegraphics[width=1\linewidth]{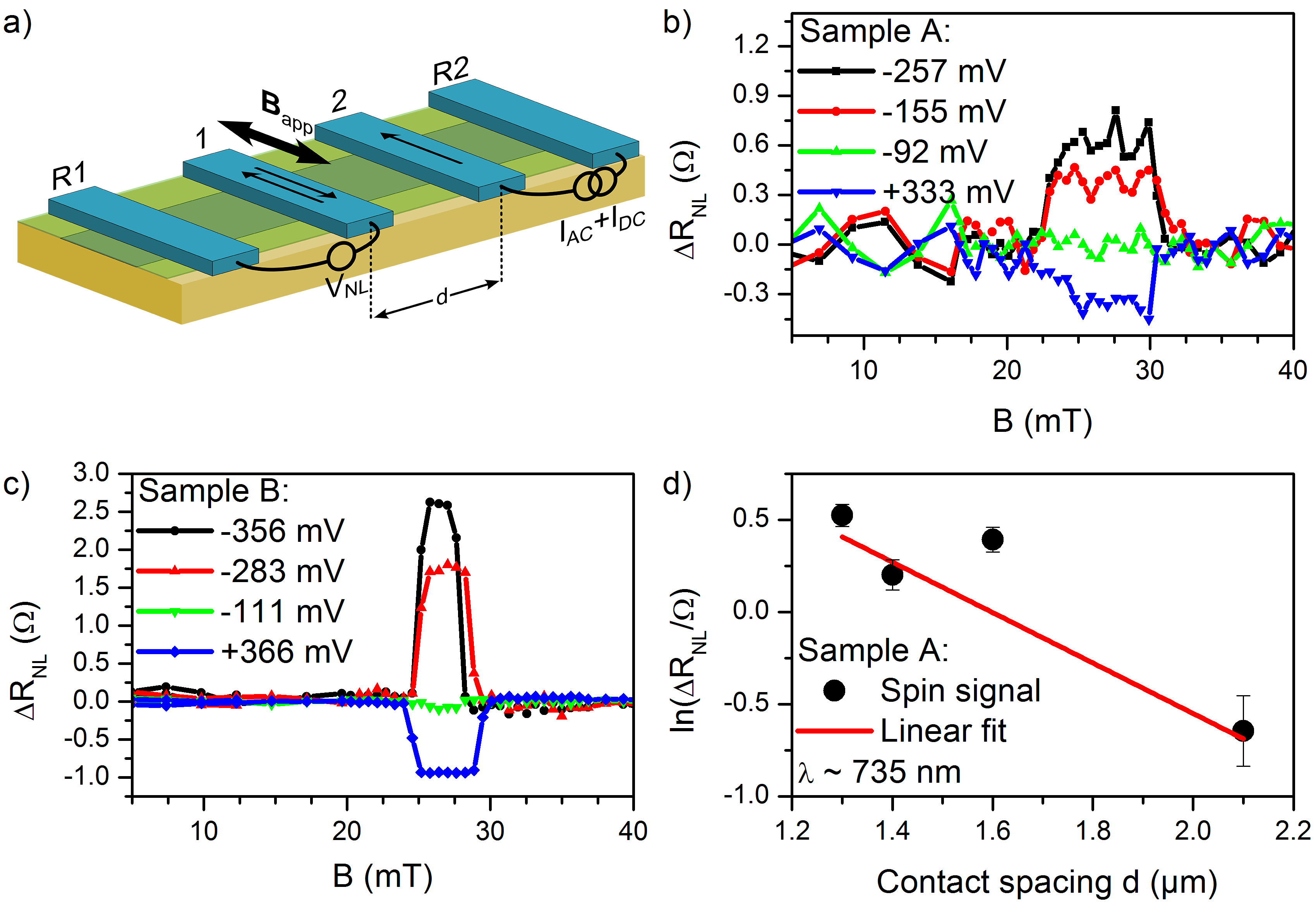}}
\caption{a) Schematic setup for a non-local spin valve measurement. b) Non-local spin valve measurements of sample~A (bl-hBN/graphene). The size of the switch between parallel and antiparallel states of contacts~1 and 2 can be tuned with the applied DC bias and is shown for four different values. c) Sample~B (bl-hBN/bl-graphene) shows a comparable dependence on the applied DC bias. Note that the spin signal changes the sign around -92~mV. d) The distance dependent spin valve measurements of sample~A allow the estimation of the spin relaxation length from the slope of the linear fit. The same analysis for sample~B is discussed in the supplementary material. \label{Fig2}}
\end{figure}

We now discuss the spin transport in graphene on YIG with a bl-hBN tunnel barrier in a non-local geometry (Fig.~\ref{Fig2}a). A current of I$_\mathrm{AC}$ = 1~$\mu$A is sourced and modulated with 3.7~Hz between contacts~2 and R2. The ferromagnetic electrode injects a spin current into the graphene underneath contact~2. These spins are diffusing along the graphene channel and are probed by a lock-in as a voltage difference $\mathrm{V}_\mathrm{NL}$ between the detector contact~1 and the reference electrode R1. Using this technique, we can decouple charge and spin transport. The signal can be defined as non-local resistance and calculated via $\rnl = \mathrm{V}_\mathrm{NL}/\mathrm{I}_\mathrm{AC}$.
To characterize the basic spin transport properties of the samples an in-plane magnetic field parallel to the electrodes (B$_\mathrm{app}$) is applied to switch the magnetization of the injector and detector (Fig.~\ref{Fig2}a). Depending on the relative magnetization alignment of the injector and detector electrodes, the non-local resistance changes between the parallel and the antiparallel resistance states when the contact magnetization switches. This measurement represents a characteristic spin valve behavior (Figs.~\ref{Fig2}b and \ref{Fig2}c) and gives an estimation of the spin relaxation length in the graphene flake (Fig.~\ref{Fig2}d).

To study the effect of the bias on the spin injection, we apply a DC current additionally to the AC current sourced between injector and reference electrode (Fig.~\ref{Fig2}a). The dielectric strength of hBN is approximately 1.2~V/nm \cite{Hattori2014}. Therefore, we limit the DC bias current for sample~A to 20~$\mu$A, which corresponds to 0.4 -- 0.6~V, depending on the IV characteristics of the injector contact. To compare different contacts, we calculate the equivalent voltage $\vhbn$ across the hBN tunnel barrier from the applied DC bias current and discuss all results plotted as function of $\vhbn$.

Figure~\ref{Fig2}b contains the spin valve measurements of sample~A for four different DC bias currents over distance d = 1.6~$\mu$m. While no spin signal above noise level is visible at -92~mV, a DC bias current of +333~mV results in a clear switching between parallel and antiparallel states with a spin signal of approximately 0.4~$\ohm$. Beyond -92~mV, we find an inverted sign of the non-local resistance switching and a spin signal of -0.4~$\ohm$ at -155~mV and -0.7~$\ohm$ at -257~mV. 

Four spin valve measurements of sample~B are shown in Fig.~\ref{Fig2}c. where we find compared to sample~A a larger spin signal of up to -2.5~$\ohm$ at -356~mV DC bias. The change of the sign of the spin signal occurs in sample~B also between -100~mV and 0~mV, a similar range as in the measurements on sample~A.

The distance dependence of the spin signal is shown for sample~A in Fig.~\ref{Fig2}d, from which we extract the spin relaxation length $\lambda \sim$ (740 $\pm$ 570)~nm.
In our previous work we found a comparable value of $\lambda = (490 \pm 40)$~nm for a not hBN protected sample. We conclude that even though the charge transport properties have improved significantly, the spin transport parameters remain similar. The same analysis was applied to sample~B, where we found $\lambda \sim$ (2.3 $\pm$ 1)~$\mu$m (supplementary material). 

The bl-hBN tunnel barriers in Fig.~\ref{Fig2}d show a less clear trend in the distance dependence, resulting in a larger error in $\lambda$. We can attribute this to two origins: an inhomogeneity of the bl-hBN tunnel barriers and an inhomogeneity in the graphene flake. Microscopic cracks in the hBN tunnel barrier could arise during the fabrication and could lead a to a different spin polarization of each contact. This interpretation is also supported by the considerable spread of the resistance-area product of between 5 to 30~k$\ohm\mu$m$^2$. As a consequence, the values for the spin relaxation length extracted from the distance dependent measurements can only be seen as approximation. However, the consistency with the spin precession measurements as discussed in the following sections confirms the validity of the estimation.

\begin{figure}[htb]
\centerline{\includegraphics[width=1\linewidth]{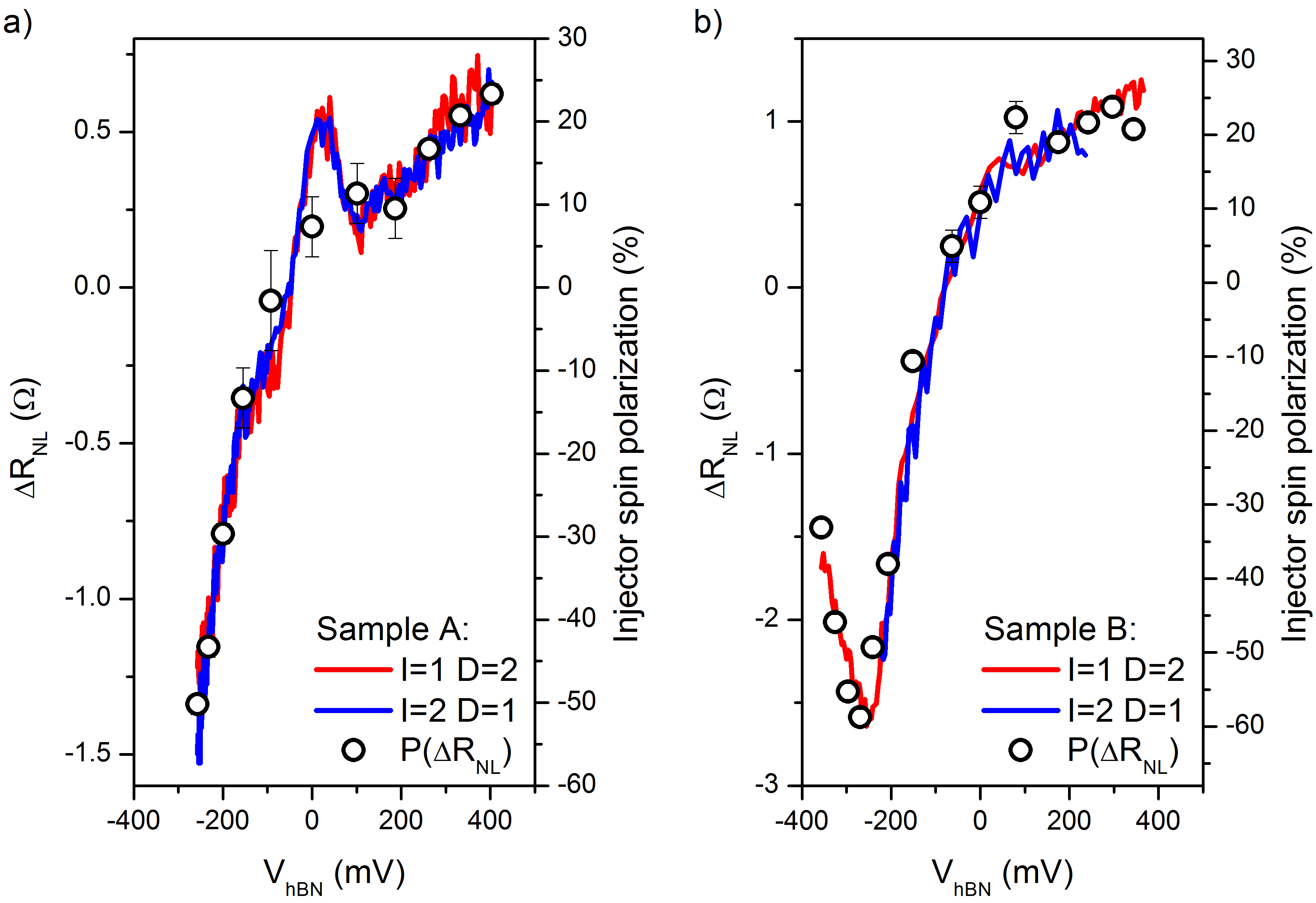}}
\caption{Non-local spin transport in a) sample~A and b) sample~B for different DC bias voltages. For comparison the dependence is shown as a function of the bias voltage applied across the hBN barrier. The blue and red curves correspond to the configuration where detector and biased injector contacts are swapped. The spin polarization on the right side of both panels is extracted from the independently measured $\Delta \rnl$. 
 \label{Fig3}}
\end{figure}

To extract the DC bias dependence of the spin injection polarization in the cobalt/bl-hBN/graphene/YIG system, we align the magnetization of injector and detector parallel or antiparallel and sweep the DC bias current. $\Delta \rnl$ = $\rnl$(P) - $\rnl$(AP) is calculated and yields the pure spin signal of samples~A and B shown in Figs.~\ref{Fig3}a and \ref{Fig3}b. For comparison, both curves are plotted as a function of $\vhbn$. 
While both positive and negative DC biases lead to an enhanced spin injection, a sign change at approximately -80~mV is observed. To extract the bias dependence of the spin injection polarization, we use the unbiased non-local spin signals to calculate the average spin polarization ($\sqrt{\pin\pd}$) of injector $\pin$ and detector $\pd$. This assumption is justified by the similar shape of the non-local resistances in Figs.~\ref{Fig3}a and \ref{Fig3}b, when injector and detector contacts are swapped. This suggests a similar behavior of both contacts. We can extract a spin polarization via:
\begin{align}
\pin \cdot \pd = \frac{\Delta \rnl \cdot \mathrm{w}}{\rsq \cdot \lambda} e^{-\mathrm{d}/\lambda} 
\end{align}
where $\Delta \rnl$ the spin signal, w the width of the flake, $\rsq$ the square resistance, $\lambda$ the spin relaxation length and d the injector to detector distance measured from the centers. Under the assumption that $\pin = \pd$ we obtain an unbiased spin polarization of 14.65\% for sample~A and 10.86\% for sample~B. Because we apply the DC bias only to the injector contact, the spin polarization of the detector remains constant and can be used to extract the dependence of the differential spin injection polarization on the DC bias. We note that the feature of sample~A around zero DC bias seems to be a characteristic feature of these particular contacts and does not appear on all contacts on sample~A (see supplementary information). 

\section{Bias dependent spin precession measurements and estimation of the proximity induced exchange field in bl-hBN/graphene/YIG}
\begin{figure}[htb]
\centerline{\includegraphics[width=1\linewidth]{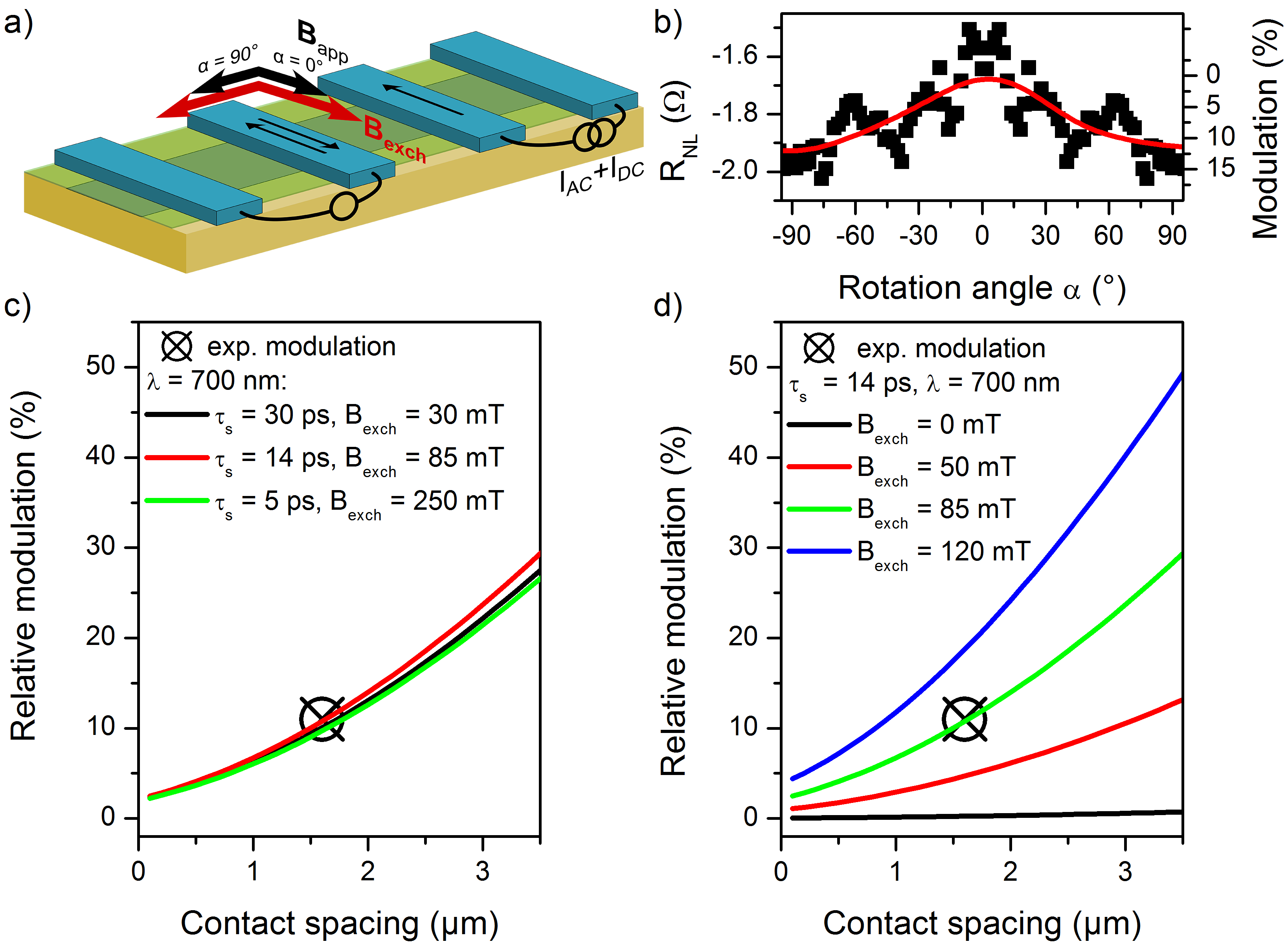}}
\caption{Modulation of spin transport with the exchange field in sample~A. a) Schematics of the experiment. $\bapp$ is rotating the YIG magnetization and the exchange field $\bexch$ in the sample plane while leaving the electrodes and injected spins unaffected. b) The angle dependence of the non-local resistance is measured at T = 10~K and -20~$\mu$A DC bias in parallel and antiparallel alignment. The subtracted spin signal is symmetrized. As a guide to the eye the smoothed data is shown in red, from which we estimate a relative modulation of 11\%. c) Fitting of the experimental relative modulation of 11\% with our model using $\tau_\mathrm{s}$ and $\bexch$ as fitting parameters. $\lambda$ = 700~nm and $\bapp$ = 15~mT are fixed parameters. d) Relative modulation of the spin signal calculated from the model using best fit parameters $\tau_\mathrm{s}$ = 14~ps and $\lambda$ = 700~nm, obtained as shown in Fig.~\ref{Fig5}. $\bexch$ is varied as indicated, and $\bapp$ = 15~mT.
 \label{Fig4}}
\end{figure}

To estimate the strength of the induced exchange field, we apply and rotate a small magnetic field ($\bapp$ = 15~mT) in the sample plane (Fig.~\ref{Fig4}a). The low in-plane coercive field of the YIG films allows us to rotate the YIG magnetization and simultaneously the proximity induced exchange field while leaving the magnetization of the cobalt injector and detector remain unaffected. The resulting modulation of the non-local resistance is a direct consequence of $\bapp + \bexch$ and can be only explained by the presence of such \cite{Leutenantsmeyer2017,Singh2017}. 

The analysis of this effect gives us an estimate for the strength of the exchange field and allows us the fitting of the Hanle curves to extract further spin transport parameters. The higher order oscillations that remain in the symmetrized data in Fig.~\ref{Fig4}b could indicate the presence of local stray fields of the cobalt contacts influencing the local YIG magnetization or an anisotropy arising from the shape of the YIG substrate which might not be fully aligned with the applied magnetic field of 15~mT. Therefore, we apply a smoothing on the data. The resulting curve is shown in red. We estimate the modulation to be (11 $\pm$ 5)\% over d = 1.6~$\mu$m, which, given the uncertainty arising from the smoothing process, should be seen as a rather rough approximation. Despite the uncertainty of the exact value of the modulation, the angular dependence indicates the presence of an exchange field in the sample. 

\begin{figure}[htb]
\centerline{\includegraphics[width=1\linewidth]{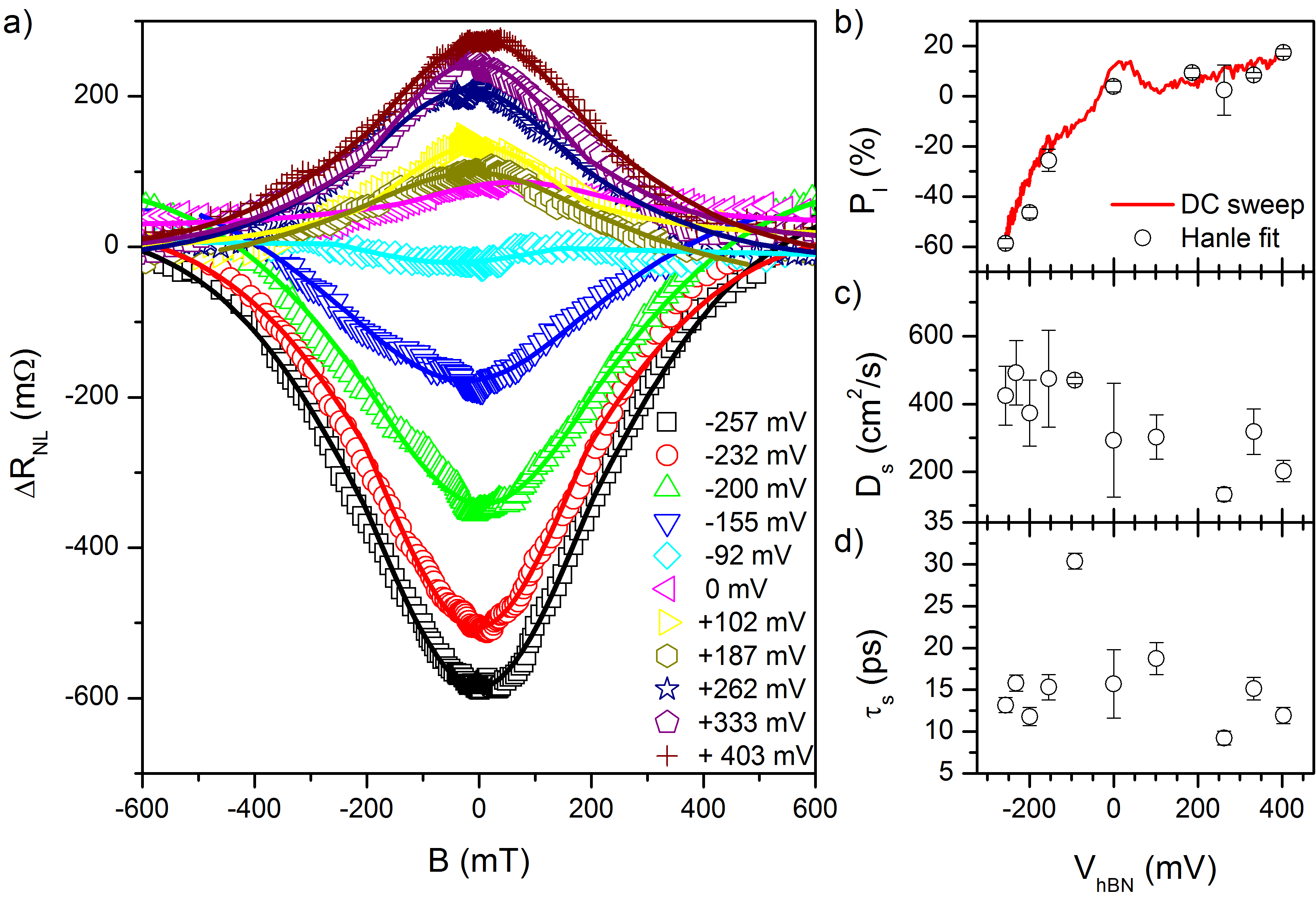}}
\caption{Spin precession measurements in sample~A: a) The Hanle spin precession curves from sample~A are fit using our exchange model with $\bexch$ = 85~mT (solid lines) for different DC bias currents. Contact~1 is used as injector, contact~2 as detector (Fig.~\ref{Fig2}a). We extract b) the calculated spin polarization the injector ($\pin$), c) the spin diffusion coefficient $\ds$ and d) the spin diffusion time $\ts$. The DC bias dependence $\pin$ shows a similar dependence as (red line in panel b, Fig.~\ref{Fig3}d).
 \label{Fig5}}
\end{figure}

Using the model reported in Leutenantsmeyer et al. \cite{Leutenantsmeyer2017} we can simulate the modulation of a spin current by exchange field induced precession. To estimate the magnitude of the exchange field leading to 11\% modulation, we use $\lambda$ = 700~nm (Fig.~\ref{Fig2}d) and assume $\ts$ to be between 5 and 30~ps, a common range for our single layer graphene devices on YIG. 
To match the experimental modulation, an exchange field between 0 and 250~mT is required (Fig.~\ref{Fig4}c). To determine the exact value of $\ts$, we use the parameter pairs of $\ts$ and $\bexch$ to fit, as discussed later, the spin precession measurements in Fig.~\ref{Fig5}a. By comparing both, we find that the both measurement sets can only be fit consistently with $\ts$ = 14~ps and $\bexch$ = 85~mT. 

Fig.~\ref{Fig4}d contains the modulation caused by the combination of the applied magnetic field of 15~mT and different values for the exchange field. The expected relative modulation caused by an applied magnetic field of 15~mT with $\lambda$ = 700~nm and $\tau_\mathrm{s}$ = 14~ps does not exceed 0.5\%, whereas the observed modulation is clearly larger. To fit the experimentally found modulation of 11\%, we have to assume $\bexch$ = 85~mT. This is a strong indication for the presence of an exchange field in this device. 
We can conclude that within the uncertainty range of the relative modulation of (11 $\pm$ 5)\%, the exchange field in sample~A is (85 $\pm$ 35)~mT.

The Hanle measurements are carried out in parallel and antiparallel alignment of the injector (contact~1) and detector (contact~2), see Fig.~\ref{Fig2}a for the contact labeling. We extract the spin signal by calculating $[\rnl(\mathrm{P})-\rnl(\mathrm{AP})]/2$, shown in Fig.~\ref{Fig5}a. From the Hanle fit using an exchange field of 85~mT, we extract the polarization of the injector P (Fig.~\ref{Fig5}b), the spin diffusion coefficient $\ds$ (Fig.~\ref{Fig5}c) and the spin diffusion time $\tau_\mathrm{s}$ (Fig.~\ref{Fig5}d). While $\ds$ = (350 $\pm$ 65)~cm$^2$/s and $\tau_\mathrm{s}$ = (16 $\pm$ 5)~ps remain approximately constant over the applied DC bias range we find a dependence of the injector spin polarization that resembles the DC bias dependence of the injector (Fig.~\ref{Fig3}a), which implies a consistency in the analysis. 
Using the spin diffusion coefficient $\ds$ and time $\tau_\mathrm{s}$ extracted from the Hanle measurements, we can calculate the spin relaxation length $\lambda = \sqrt{\ds \tau_\mathrm{s}} = (730 \pm 230)$~nm. When compared to the estimation from the distance dependent spin valve measurements (Fig.~\ref{Fig2}a) both approaches yield similar values which indicates again the consistency of the analysis. 

Note that the rather smooth Hanle curves shown in Fig.~\ref{Fig5}a could be also fit with a conventional spin precession model that does not include any exchange field. These fittings yield $\ts$ $\sim$ 25~ps, $\ds$ $\sim$ 800~cm$^2$/s and $\lambda$ $\sim$ 1.4~$\mu$m. Apart from $\ds$ being unrealistically large, the extracted $\lambda$ is two times larger than the result from the independently measured distance dependent spin valves (Fig.~\ref{Fig2}d) which suggests that the fit of our results with the conventional model is unreliable. 
Furthermore, if we want to fit the modulation in Fig.~\ref{Fig4}b with $\lambda$ = 1.4~$\mu$m and $\ts$ = 25~ps, an exchange field of $\sim$ 60~mT would be required to match the data, even though the Hanle fitting did not include any $\bexch$. In return, the parameter sets that match 11\% modulation do not fit the spin precession measurements unless the values are close to $\lambda$ = 700~nm, $\ts$ = 14~ps and $\bexch$ = 85~mT. 
In conclusion, this analysis underlines the relevance to carry out both, angular modulation of $\rnl$ and Hanle precession experiments, to characterize the exchange field strength.

\section{Bias dependent spin precession measurements in bl-hBN/bl-graphene/YIG}
\begin{figure}[htb]
\centerline{\includegraphics[width=1\linewidth]{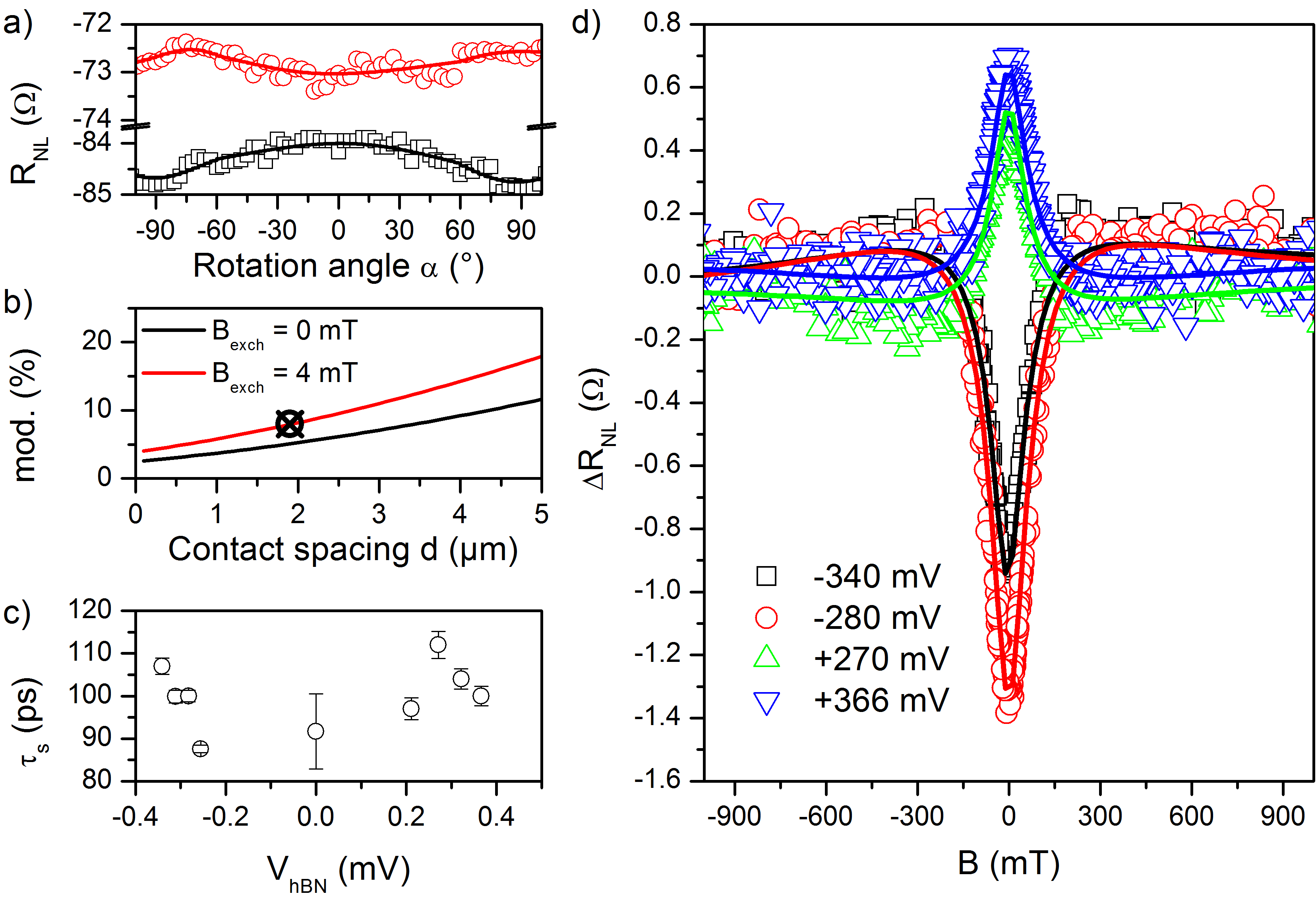}}
\caption{a) The non-local resistance can be modulated by 8\% by rotating an in-plane magnetic field of 15~mT. The solid lines are smoothed and a guide to the eye. The red line is measured in parallel alignment, the black line in antiparallel configuration. b) Modeling of the 8\% modulation with the spin transport parameters of $\lambda$ = 2.3~$\mu$m and $\tau_\mathrm{s}$ = 100~ps. The black curve represents the modulation by the applied magnetic field of 15~mT in the absence of an exchange field, the red curve adds an exchange field of 4~mT. c) The spin relaxation time $\tau_\mathrm{s}$ extracted from the Hanle data in panel d. d) The Hanle spin precession curves of sample~B with the fitting curves (lines) for different DC bias currents. The spin relaxation length of $\lambda$ = 2.3~$\mu$m is used as parameters for the fitting.
 \label{Fig6}}
\end{figure}

In comparison to sample~A, sample~B is fabricated with a bilayer graphene flake. The extraction of the spin relaxation length via distance dependent spin valve measurements is done in a similar way as for sample~A and is shown in the supplementary information in Fig.~\ref{FigS4}. We extract $\lambda = (2.3 \pm 1)\,\mu$m. The modulation of the non-local resistance by rotating the exchange field in the sample plane is shown in Fig.~\ref{Fig6}a. The parallel (red) and antiparallel (black) data is measured at 10~K and -366~mV DC bias. The solid line is the smoothed data and used to estimate the relative modulation of the spin signal after subtraction of the parallel and antiparallel data which results in a modulation of 8\%.

To estimate the exchange field causing this precession, we use $\lambda$ = 2.3~$\mu$m extracted for sample~B from the distance dependent measurements and assume $\tau_\mathrm{s}$ = 100~ps, which is later confirmed by the Hanle spin precession measurements. In this particular case, the modulation of the applied magnetic field of 15~mT (black line, Fig.~\ref{Fig6}d) already induces a modulation close to the experimentally found one. To match the data, a very small exchange field of only 4~mT would be required, leading us to the conclusion that in this device most likely no exchange interaction is present. 

Using the Hanle spin precession data, we also extract $\lambda$ = 2.3 $\mu$m with a negligible exchange field. We find consistently over all biases a spin diffusion time of (100 $\pm$ 8)~ps and a spin diffusion coefficient of $\ds = \lambda^2/\tau_\mathrm{s}$ = (530 $\pm$ 40)~cm$^2$/s, which resembles the values used for the modulation fit and indicates consistency throughout our analysis of the spin transport. 
The possible absence of the exchange field in sample~B stresses the importance of the graphene/YIG interface of these devices. This observation could be also explained with a different proximity effect on each of the two bilayer graphene layers.
Nevertheless, sample~B shows a similar dependence on the applied DC bias as sample~A and shows that the tunable spin injection is also present in the bl-hBN/bl-graphene/YIG system.

\section{Conclusion}
We have studied the spin injection through bl-hBN tunnel barriers into single- and bilayer graphene on YIG, showing a more reliable and efficient spin injection compared to TiO$_\mathrm{x}$ tunnel barriers. The bl-hBN tunnel barriers yield a resistance-area product between 5 and 30~k$\ohm\mu$m$^2$ and the spin injection polarization is found to be tunable through a DC bias current applied to the injector. We observe a sign inversion at approximately -80~mV DC bias applied across the bl-hBN flake. We estimate the proximity induced exchange field through in-plane and out-of-plane spin precession measurements to be around 85~mT in sample~A and likely to be absent in sample~B. The low magnitude of the exchange field compared to theoretical predictions emphasizes the importance of the graphene/YIG interface on the proximity induced exchange field and confirms our previously reported low exchange strength for graphene/YIG devices. Nevertheless, our results confirm the unique properties of bl-hBN for the reliable spin injection into single and bilayer graphene on YIG and stress the importance of this type of tunnel barrier for future application in graphene spintronics.

\section{Acknowledgements}
We acknowledge the fruitful discussions with J.~Ingla-Ayn\'es, 
and funding from the European Union’s Horizon 2020 research and innovation program under grant agreement No 696656 and 785219 (‘Graphene Flagship’ core~1 and 2), the Marie Curie initial training network ‘Spinograph’ (grant agreement No 607904) and the Spinoza Prize awarded to B.J. van Wees by the ‘Netherlands Organization for Scientific Research’ (NWO).

\bibliography{references}

\widetext
\clearpage
\begin{center}
\textbf{\large Supplementary Information}
\end{center}
\setcounter{equation}{0}
\setcounter{figure}{0}
\setcounter{section}{0}
\setcounter{table}{0}
\makeatletter
\renewcommand{\theequation}{S\arabic{equation}}
\renewcommand{\thesection}{S\Roman{section}}
\renewcommand{\thefigure}{S\arabic{figure}}
\renewcommand{\citenumfont}[1]{S#1}

\section{Full set of the hBN tunnel barrier characterization}
\begin{figure}[H]
\centerline{\includegraphics[width=0.7\linewidth]{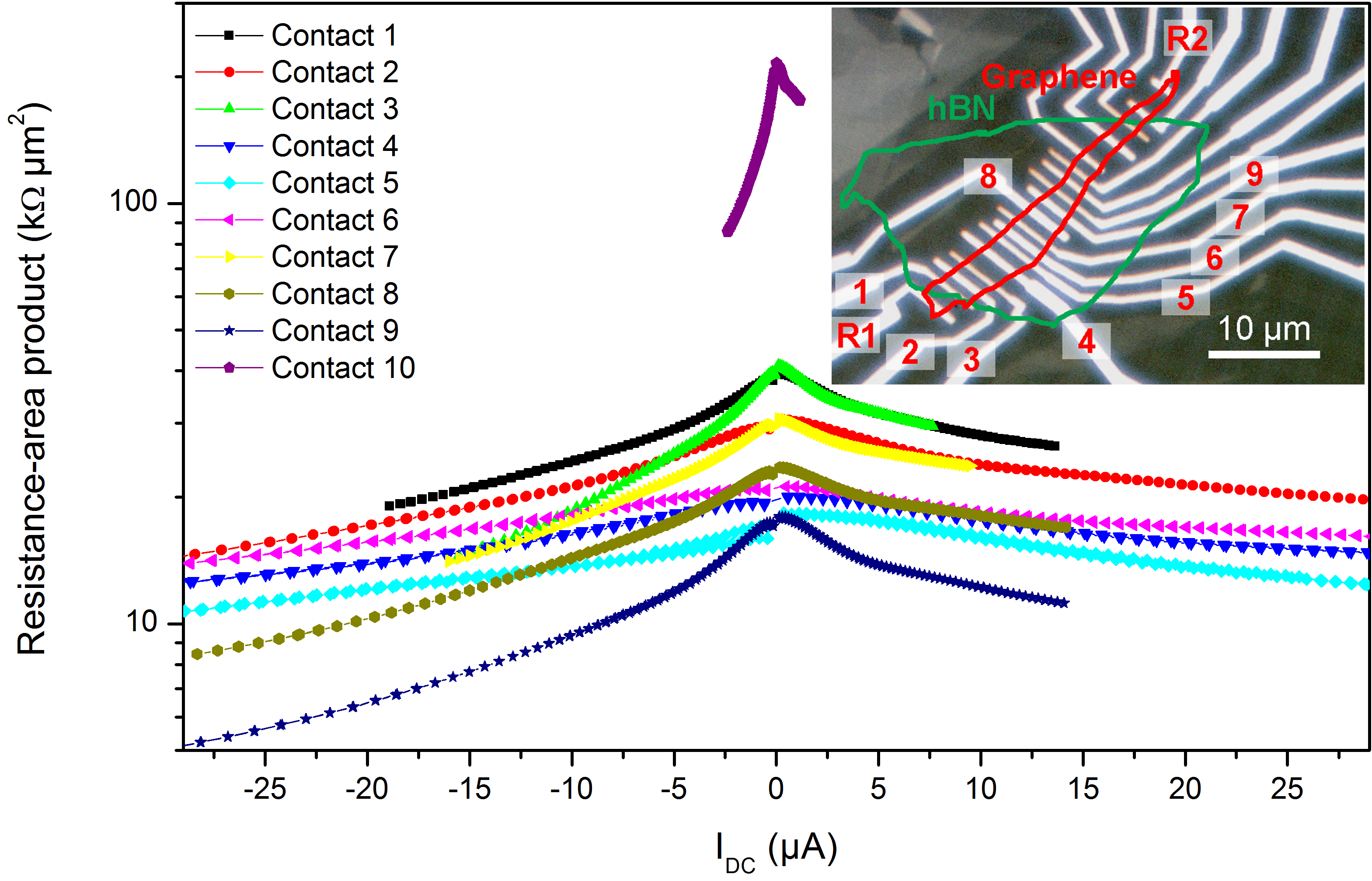}}
\caption{Full set of the contact characterization of sample~A. The inset shows the microscope image with the characterized contacts. All contacts with a bilayer hBN tunnel barrier have a relatively homogeneous resistance-area product. Given the significantly higher resistance of contact~10, we suppose that this contact has a trilayer hBN tunnel barrier. 
 \label{FigS1}}
\end{figure}

\begin{figure}[H]
\centerline{\includegraphics[width=0.7\linewidth]{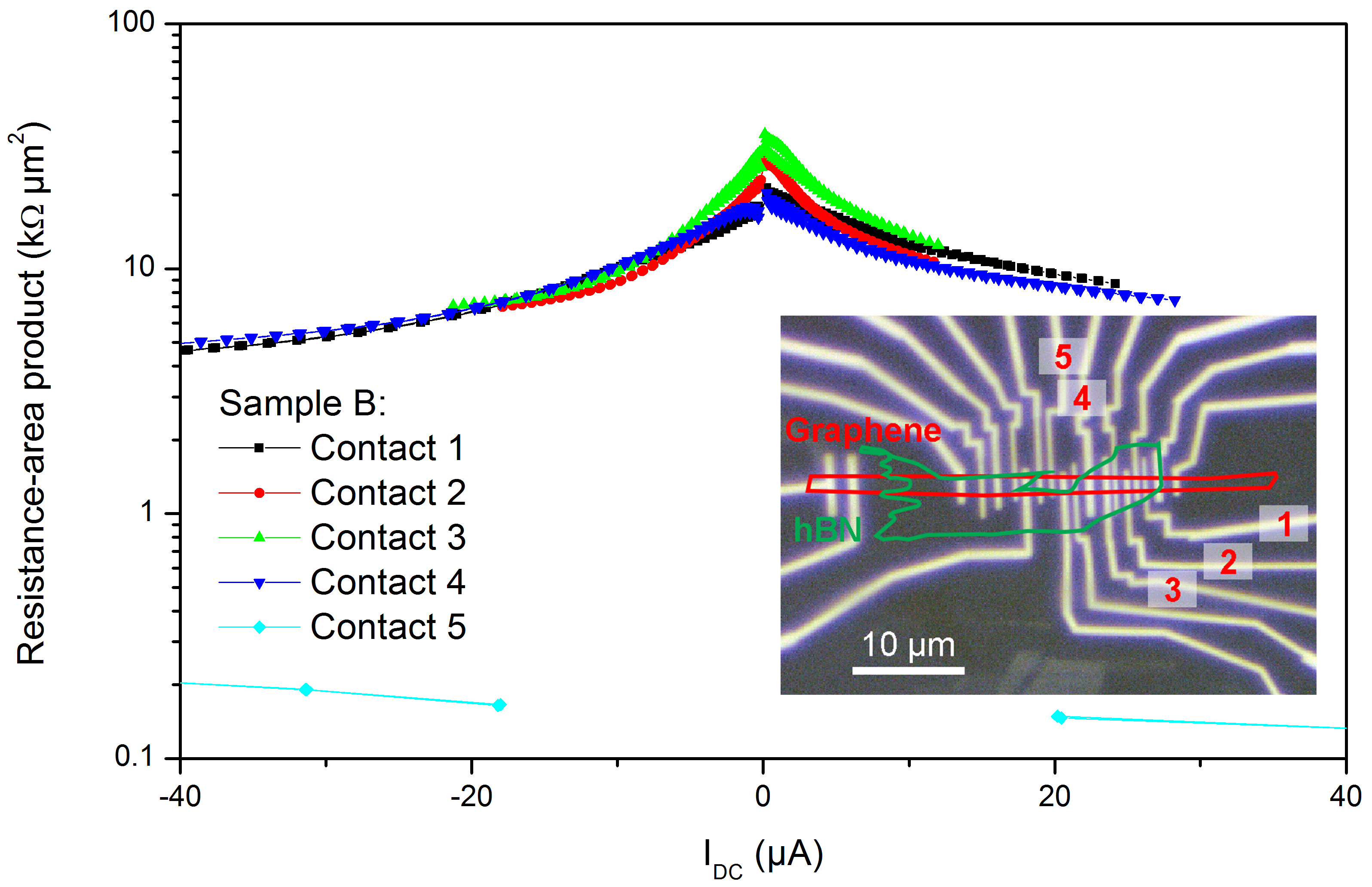}}
\caption{Extended measurements of  the contacts on sample~B. The inset shows the microscope image with the characterized contacts. Contact~5 shows a linear metallic behavior, due to the shape of the hBN flake the cobalt is likely in direct contact with the graphene flake.  
 \label{FigS2}}
\end{figure}

\begin{figure}[H]
\centerline{\includegraphics[width=0.7\linewidth]{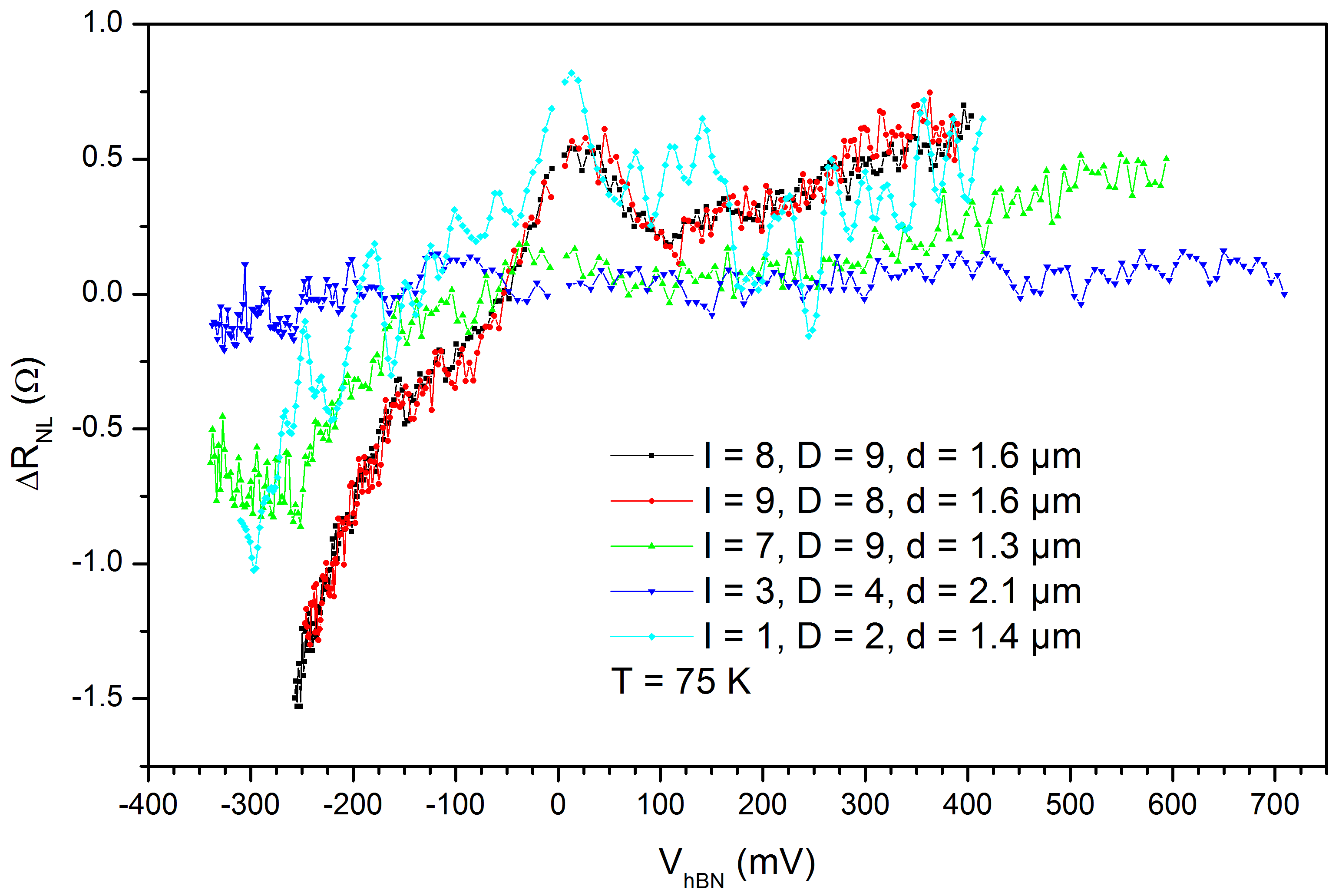}}
\caption{Extended measurements of DC bias sweeps on sample~A. See inset of Fig.~\ref{FigS1} for the contact numbering. The data is obtained by aligning the injector~I and the detector~D parallel and antiparallel and subtracting both curves. Since the detection polarization remains constant over the applied bias range, the increase of the non-local resistance corresponds to the increase of the spin injection polarization, which is relatively homogeneous over the contacts. The first two curves are discussed in the main text.
 \label{FigS3}}
\end{figure}

\section{Estimation of the spin relaxation length in sample~B}
\begin{figure}[H]
\centerline{\includegraphics[width=0.7\linewidth]{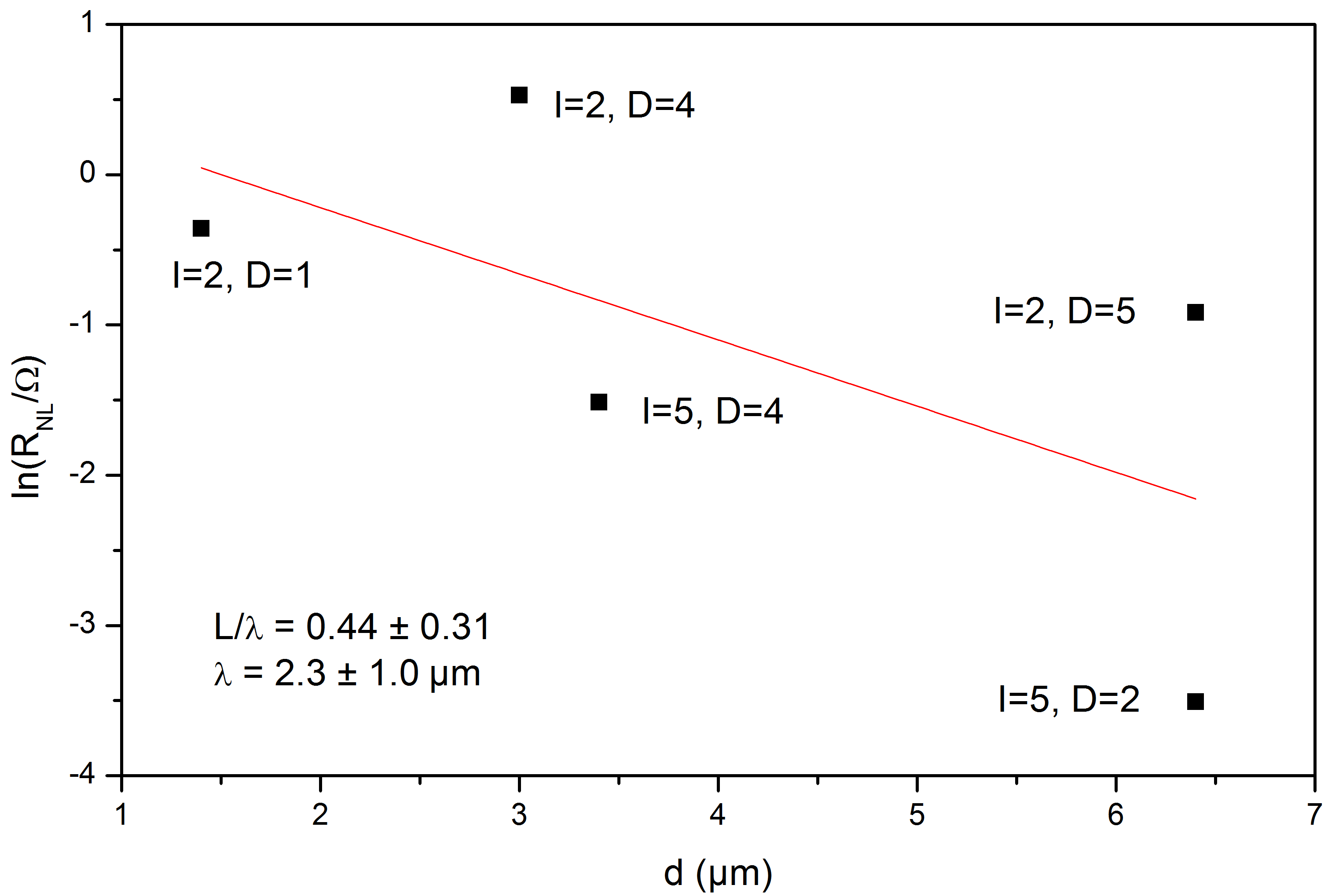}}
\caption{Distance dependent measurements of the spin valves on sample~B. The large difference in the magnitude of the spin signal indicates an inhomogeneous spin polarization of the contacts and could be caused by cracks in the bl-hBN flake. See the inset of Fig.~\ref{FigS2} for the contact numbering.
 \label{FigS4}}
\end{figure}

\section{Origin of the background of the Hanle curves in sample~B}
The data shown in the main text in Fig.~\ref{Fig6} contains only the pure spin signal between injecting and detecting electrode. The spin signal is obtained by aligning the injector and detector parallel and antiparallel and subtracting both curves. The remaining signal is in theory the purely spin dependent signal. Spurious effects that are present in the measured signal are hereby extracted. These effects can be obtained by calculating the background signal by adding the parallel and antiparallel Hanle curves. 

\begin{figure}[H]
\centerline{\includegraphics[width=0.7\linewidth]{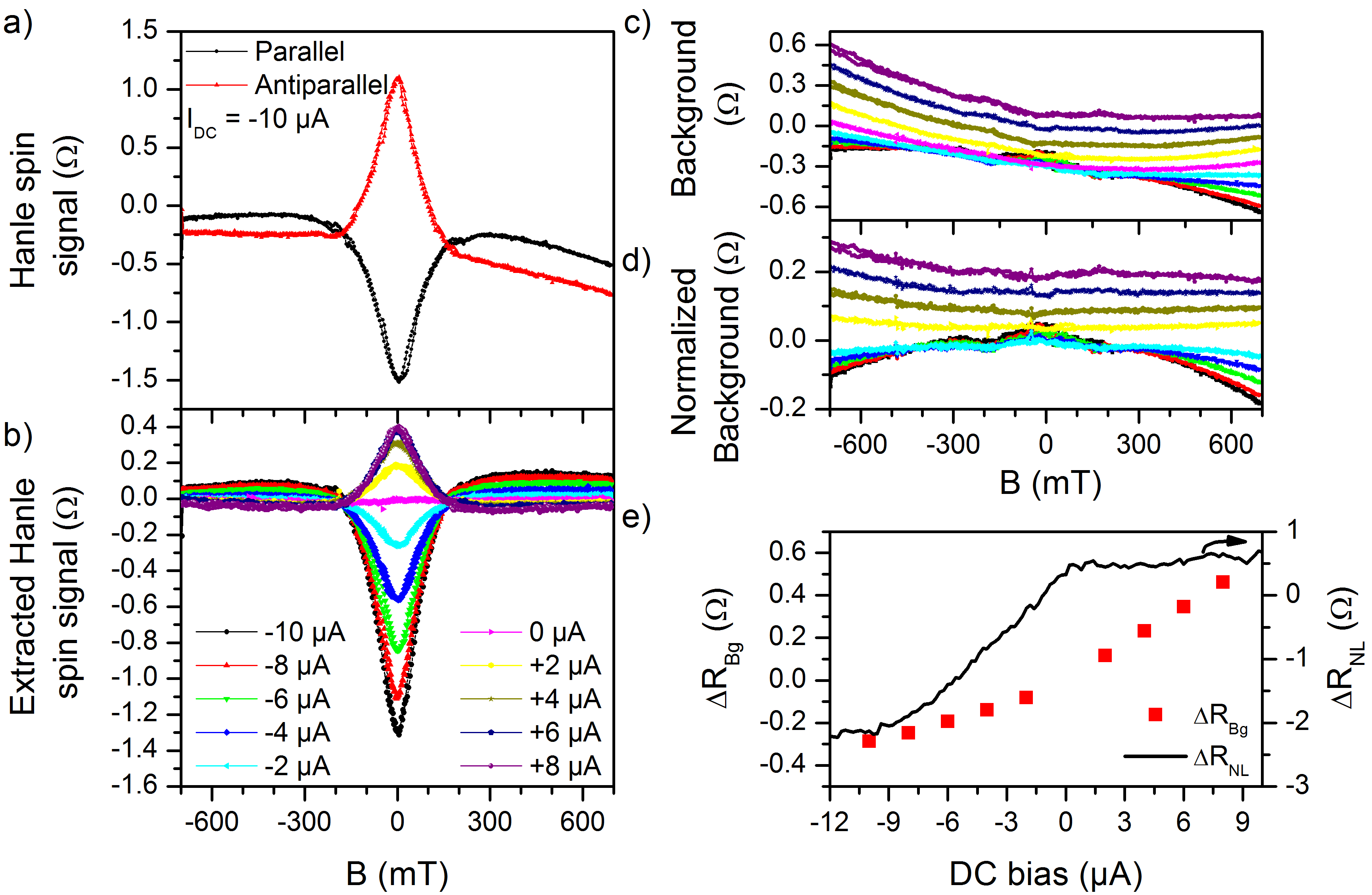}}
\caption{a) The raw data of the Hanle measurements on sample~B has a significant background signal that is excluded from b) the spin signal. The dependence of the background signal on the applied DC bias shown in panel c). The background signal  is extracted by adding the antiparallel to the parallel Hanle curve. d) To separate the spin and charge dependent contributions to the background signal, we subtract the data measured with the minimized spin signal (0~$\mu$A DC bias) from the individual Hanle background curves and extract the shown background signal. e) The amplitude of the Hanle background signal shows a dependence on the DC bias that roughly resembles the inverted dependence of the injector and detector electrode, which could indicate that the background signal has still a spin related contribution coming from one of the reference contacts.
 \label{FigS5}}
\end{figure}

In Fig.~\ref{FigS5}a we show the measured Hanle curves, the extracted spin signal in Fig.~\ref{FigS5}b and the extracted background signal in Fig.~\ref{FigS5}c. Both spin and background signal show a dependence on the applied DC bias. The presence of a spin related signal in the background signal is not expected, however, the dependence on the DC bias suggests the opposite case. 

To determine the nature of the signal, we normalize the data set to the signal where the spin signal and the spin injection polarization is minimized, which is here the case for a DC bias of 0~$\mu$A (Fig.~\ref{FigS5}b). This way we can separate the charge and spin dependent signals in the background data that do not depend on the magnetization of the inner detector and injector electrodes. The resulting signal is shown in Fig.S5c. We find a clear dependence on the applied DC bias. We suspect this signal to arise either as contribution from the current reference electrode or as the rotation of the cobalt electrodes at high magnetic fields out of the sample plane.

If we compare the signal amplitude averaged at $\pm$700~mT ([$\rnl$(+700~mT)+$\rnl$(-700~mT)]/2), we find a dependence on the DC bias as shown by the red squares in Fig.~\ref{FigS5}e. This slope approximately resembles that of the DC bias measurements but of opposite sign, which suggests that this signal might be actually spin related. Since the inner injector and detector signals are excluded from this data, we can identify the injector reference contact to be likely the origin. This contact is also biased with the DC current but does not have an hBN tunnel barrier. Therefore, the observation of such large signal is still surprising, especially for of the greater distance of the reference electrode to the detector of 4~$\mu$m instead of 1.9~$\mu$m. At this moment, we are unable to determine the origin of the DC bias dependence of the background signal. Further work is needed for clarification. 

\end{document}